\documentclass{article}

\usepackage[preprint,nonatbib]{neurips_2023}

\usepackage{graphicx}
\graphicspath{ {./images/} }

\usepackage[utf8]{inputenc} 
\usepackage[T1]{fontenc}    
\usepackage{hyperref}       
\usepackage{url}            
\usepackage{booktabs}       
\usepackage{amsfonts}       
\usepackage{nicefrac}       
\usepackage{microtype}      
\usepackage{xcolor}         
\usepackage{float}
\usepackage{wrapfig}
\usepackage[numbers]{natbib}

\title{Discovery of Novel Reticular Materials for \\ Carbon Dioxide Capture using GFlowNets}
\author{Flaviu Cipcigan \\
        IBM Research - Europe \\
        \texttt{flaviu.cipcigan@ibm.com} 
        \And
        Jonathan Booth \\
        Science and Technology Facilities Council \\
        \texttt{jonathan.booth@stfc.ac.uk}
        \And
        Rodrigo Neumann Barros Ferreira \\
        IBM Research \\
        \texttt{rneumann@br.ibm.com} \\
        \And
        Carine Ribeiro dos Santos \\
        IBM Research \\ 
        Universidade Federal do Rio de Janeiro \\
        \texttt{carineribeiro@ibm.com}
        \And
        Mathias Steiner \\
        IBM Research \\
        \texttt{mathiast@br.ibm.com}
        }
\begin{document}

\maketitle

\begin{abstract}
Artificial intelligence holds promise to improve materials discovery. GFlowNets are an emerging deep learning algorithm with many applications in AI-assisted discovery. By using GFlowNets, we generate porous reticular materials, such as metal organic frameworks and covalent organic frameworks, for applications in carbon dioxide capture. We introduce a new Python package (\texttt{matgfn}) to train and sample GFlowNets. We use \texttt{matgfn} to generate the \texttt{matgfn-rm} dataset of novel and diverse reticular materials with gravimetric surface area above 5000 m\textsuperscript{2}/g. 
We calculate single- and two-component gas adsorption isotherms for the top-100 candidates in \texttt{matgfn-rm}. These candidates are novel compared to the state-of-art ARC-MOF dataset and rank in the 90\textsuperscript{th} percentile in terms of working capacity compared to the CoRE2019 dataset. We discover 15 materials outperforming all materials in CoRE2019.
\end{abstract}

\section{Introduction}

Artificial intelligence holds promise to improve the scientific method \cite{Hey2009TheFP, Agrawal2016PerspectiveMI} and to accelerate scientific discovery. Applied to materials \footnote{Here, we conceptualise materials broadly to include molecules, proteins, crystals and complex materials.}, AI unlocks vast search spaces and enables novel applications in pharmaceuticals \cite{Das2021, Crusius2023, Hammond2021, Cipcigan2020}, batteries or carbon capture \cite{mcdonagh2023machine}.

Reticular materials \cite{Yaghi2019IntroductionTR} such as metal organic frameworks (MOFs) and covalent organic frameworks (COFs) are extended periodic structures connected via strong bonds \cite{Freund202125YO}. They are synthesized by connecting building blocks known as secondary building units to form three dimensional periodic structures \cite{Kalmutzki2018}. By choosing the building blocks, the properties of a reticular material can be tuned to support many applications \cite{Yaghi2019IntroductionTR}.

Reticular materials with high gravimetric surface area are particularly useful for applications in carbon capture, since carbon dioxide molecules adsorb at the internal surface area \cite{Farha2012}. The larger the gravimetric surface area, the more gas molecules can be absorbed per gram of material.

In this work, we use GFlowNets to generate reticular materials with high gravimetric surface area for applications in carbon capture. Our key contributions are:

\begin{enumerate}
    \item The \texttt{matgfn} Phython library for training and sampling using GFlowNets.
    \item A workflow using \texttt{matgfn} to generate reticular materials using secondary building units. 
    \item The \texttt{matgfn-rm} dataset of diverse and novel reticular materials with total internal surface area higher than 5000 m\textsuperscript{2}/g. The top-100 reticular materials candidates are novel compared to the reference ARC-MOF dataset, rank in the 90th percentile in terms of working capacity compared to the CoRE201910 dataset. We discover 15 materials outperforming all materials in CoRE2019.
\end{enumerate}

\section{Background and related work}

\textbf{Generative Flow Networks} GFlowNets \cite{Bengio2021, Bengio2021GFlowNetF} are an emerging machine learning algorithm with many applications in AI-assisted materials discovery \cite{Jain2023GFlowNetsFA}. GFlowNets learn to generate composite objects objects $\underline{x}$ sampling from an unormalised distribution $p(\underline{x}) \propto R(\underline{x})$ where $R(\underline{x})$ is a user-specified positive reward function. A composite object $\underline{x}$ consists of symbols drawn from a vocabulary $\mathbb{V}$ and relationships between those symbols. For example, $\underline{x}$ can be a sequence $\underline{x} = \left[x_1, x_2, \ldots x_n\right]$ or a graph. The object $\underline{x}$ is built by through Markov Decision Process restricted to a directed acyclic graph. 
Transition probabilities $p(x_{i+1} \, | \, \underline{x})$ are approximated by a neural network called a flow model. GFlowNets need fewer evaluations of the reward function to generate samples with high reward, novelty and diversity when compared to alternatives such as Markov Chain Monte Carlo, Proximal Policy Optimisation or Bayesian Optimisation \cite{Bengio2021}. 

\textbf{Building hypothetical reticular frameworks} Trillions of hypothetical frameworks such as MOFs or COFs can be generated by placing secondary building units \cite{Kalmutzki2018} into nodes and edges of a three dimensional topology \cite{OKeeffe2008}. A secondary building unit is an organic molecule or a coordination compound (a metal linked to organic atoms). A topology is a three dimensional arrangement of nodes and edges. Replacing nodes and edges with secondary building units results in a three dimensional point cloud of atoms connected by covalent or metal-organic bonds. We use the \texttt{pormake} secondary building units \cite{Lee2021} and topology codes from the Reticular Chemistry Structure Resource \cite{OKeeffe2008}. Previously, deep autoencoders \cite{Yao2021} and evolutionary methods \cite{Lee2021} have been used to generate frameworks using this approach. 

\textbf{Reference datasets} We use two reference datasets in this work. These datasets are not used for training models, but as comparison once training is done, as GFlowNet generates candidates using just a reward function. The CoRE2019 dataset \cite{Chung2019} consists of 12,023 metal-organic frameworks with carbon dioxide uptake properties calculated by Moosavi \textit{et al.} \cite{Moosavi2020} using Grand Canonical Monte Carlo. ARC-MOF (reported in 2022) \cite{Burner2023} is a collection of 279,610 MOFs from previous MOF datasets. It contains both experimental and hypothetical MOFs.

\section{Generating reticular frameworks with GFlowNets}

\begin{wrapfigure}{r}{0.45\textwidth}
  \begin{center}
    \includegraphics[width=0.40\textwidth]{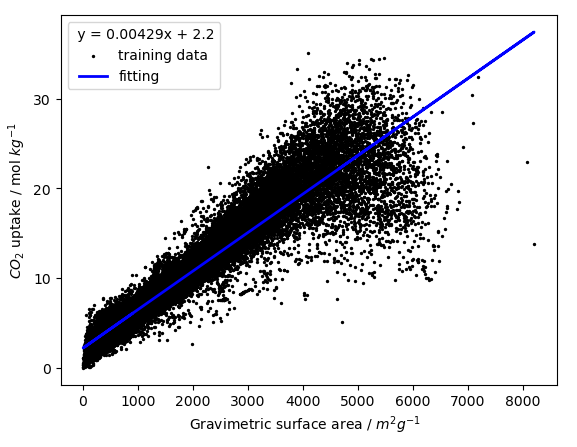}
  \end{center}
  \caption{Regression of simulated high pressure CO\textsubscript{2} uptake to gravimetric surface area}
    \label{fig:nanoporous_correlation}
\end{wrapfigure}

\textbf{Python package} We built a Python library called \texttt{matgfn} to train and sample GFlowNets. The library is built on top of PyTorch \cite{pytorch} and Gymnasium \cite{towers_gymnasium_2023} and prioritises ease of use and code readability. We intend for \texttt{matgfn} to be a general Python package for generation of diverse types of materials from small molecules to framework materials. Architecturally, \texttt{matgfn} separates sampling, loss calculation, optimisation, and environment definition as modular Python classes. Each can be modified individually, to implement off-policy training or use improved losses, for example. We note similar architectural choices for \texttt{torchgfn}.

\textbf{Environment for reticular framework generation} We configure a GFlowNet environment to build string sequences of the form \texttt{["N577", "N238", "N194", "E5", "E3", "E74"]}. Here \texttt{N} represents node building blocks and \texttt{E} represents edge building blocks in the \texttt{pormake} database. The string sequences are then transformed to a Crystallographic Information File (\texttt{*.cif}) by \texttt{pormake} to create a reticular framework. Not all strings create valid materials, so during generation, building blocks were restricted such that (a) each topology had the correct number of nodes and edges, (b) the building blocks were placed in the correct order and (c) each slot had a compatible building block.

\textbf{Reward} We calculate the Gravimetric Surface Area $GSA$ in m\textsuperscript{2}/g with Zeo++ \cite{Zeo} during the training loop of the GFlowNet. We configure Zeo++ with a probe radius of $1.525\,$\AA{} and 2000 samples. The GFlowNet is given the following reward:
\begin{equation}
    R(\underline{x}) = \mathcal{H}\left(GSA(\underline{x}) - C\right) * \exp\left({\frac{GSA(\underline{x}) - C}{C}}\right)
    \label{reward-function}
\end{equation}
where $\mathcal{H}(x)$ is the Heaviside step function $\mathcal{H}(x) = 0 \, \mbox{if } x < 0, 1  \, \mbox{if } x \geq 0$ and $C$ is a cutoff. Zeo++  and \texttt{pormake}  sometimes raise errors due to large distances between atoms. The reward is zero when an error occurred to encourage the GFlowNet to avoid materials with unrealistic bond lengths.

\textbf{Relationship with CO\textsubscript{2} capture } We check whether the gravimetric surface area predicts CO\textsubscript{2} uptake by analysing approximately 30,000 MOFs from three databases: CoRE2019 \cite{CORE_2019}, ARABG \cite{ARABG} and BW20K \cite{Boyd2019}. We performed univariate linear regression of CO\textsubscript{2} uptake at 16 bar using each of the geometric and chemical descriptors. The best performing descriptor was the gravimetric surface area with coefficient of determination is 0.88, RMSE is 2.41 mol kg\textsuperscript{-1} and Spearman's rank correlation coefficient of 0.97. Figure \ref{fig:nanoporous_correlation} shows the CO\textsubscript{2} uptake as a function of gravimetric surface area. We validated the regression using 50 rounds of 10-fold cross validation, with each cross-validation consisting of an 80-20 split between training and test data. The mean coefficient of determination is 0.88 ± 0.0002 and mean RMSE is 2.41 ± 0.022 mol kg$^-1$. The training and test values of coefficient of determination and RMSE are the same to two decimal places and the standard deviation of these metrics during cross validation are very small which shows that the correlation is robust and stable.

\section{The \texttt{matgfn-rm} dataset}
\textbf{Training} We trained a GFlowNet using Trajectory Balance loss \cite{malkin2022trajectory} and an LSTM flow model. We use a learning rate of $5 \times 10^{-3}$ for both the flow model and the partition function. We train for a maximum of 100,000 episodes and stop when the mean loss over 10,000 episodes is lower than 1.8. Eleven topologies were chosen: CDZ-E, CLD-E, EFT, FFC, TSG, TFF, ASC, DMG, DNQ, FSO, URJ. For each topology, two GFlowNets were trained, one with edges and one without. The performance is shown in Supplementary Information. Once the GFlowNets have been trained, they were sampled to generate \texttt{matgfn-rm} dataset of over 1 million hypothetical reticular frameworks.

\begin{figure}[ht]
  \begin{minipage}[b]{0.42\linewidth}
    \centering
    \includegraphics[width=\linewidth]{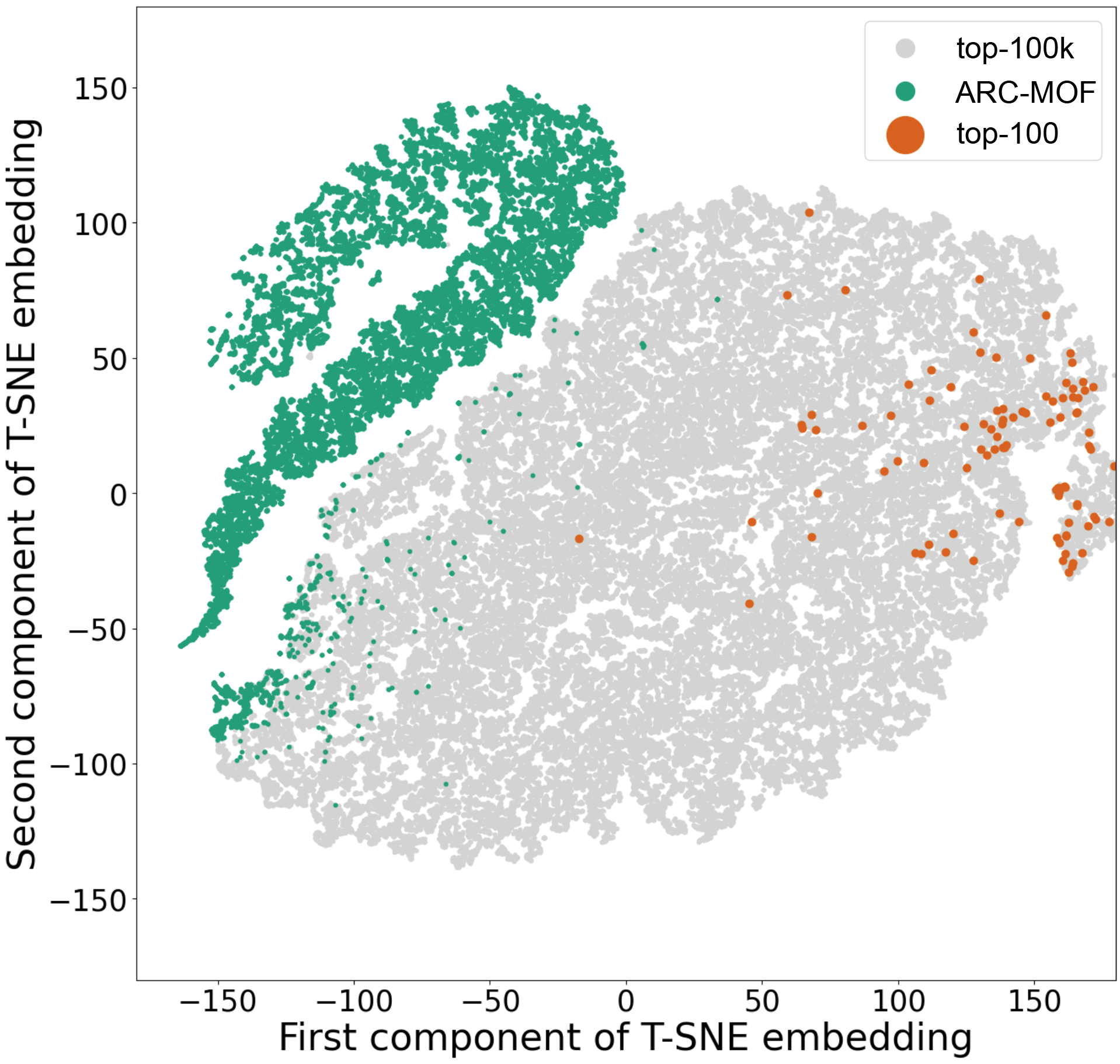}
    \caption{Two dimensional T-SNE embedding of the average minimum distance of ARC-MOF (green), the top-100,000 (gray) and top-100 (orange) materials from \texttt{matgfn-rm}.}
    \label{fig:diversity-analysis}
  \end{minipage}
  \hfill
  \begin{minipage}[b]{0.45\linewidth}
    \centering
    \includegraphics[width=\linewidth]{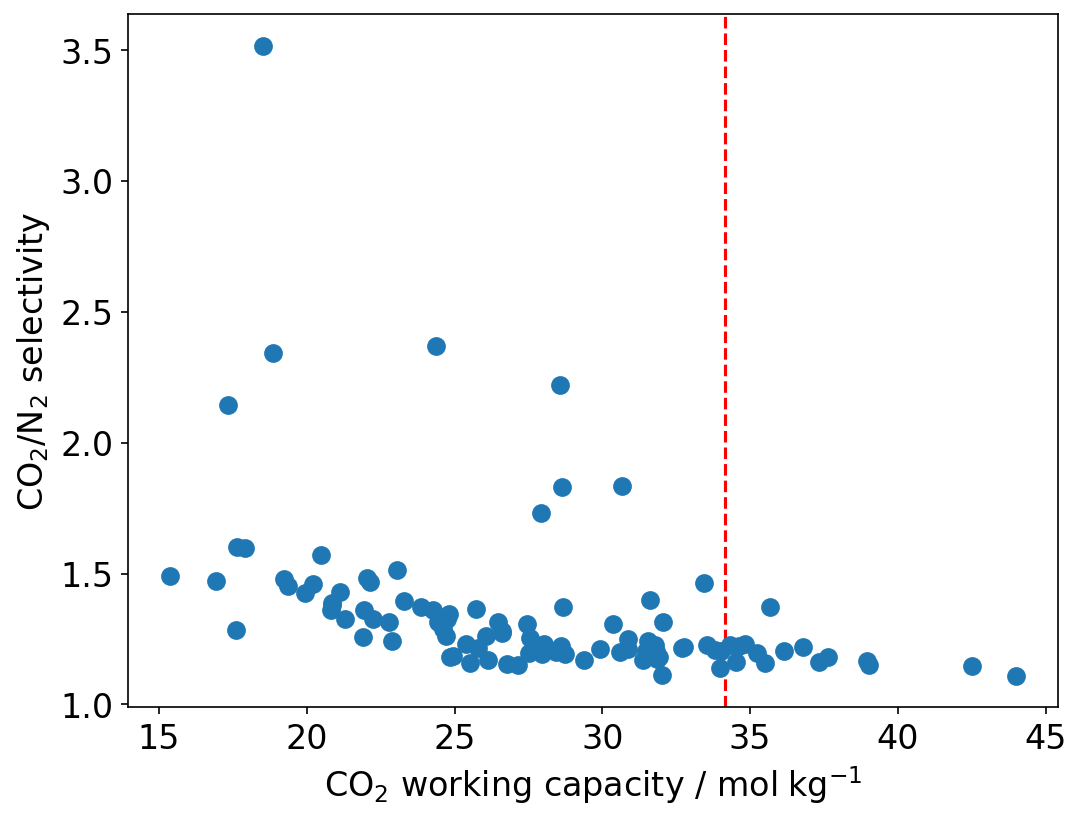}
    \caption{Simulated CO\textsubscript{2} working capacity and CO\textsubscript{2}\,/\,N\textsubscript{2} selectivity for the top-100 \texttt{matgfn-rm} materials. The (red) dashed line represents the highest working capacity found in the CoRE2019 dataset, which is surpassed by 15 of the top-100 \texttt{matgfn-rm} materials.}
    \label{fig:adsorption-simulations}
  \end{minipage}
\end{figure}

\textbf{Diversity analysis} We compare the top-100 and top-100,000 candidates from \texttt{matgfn-rm} to the ARC-MOF dataset. For each CIF file, we compute the average minimum distance (AMD) descriptor \cite{Widdowson2021} of length 100. We then perform dimensionality reduction to two dimensions using t-SNE implemented in 
\texttt{scikit-learn} \cite{scikit-learn}. Figure~\ref{fig:diversity-analysis} shows the result. There, the top-100 and top-100,000 \texttt{matgfn-rm} materials are separated from most materials from ARC-MOF. This shows that we discover new materials compared to existing datasets.

\begin{wrapfigure}{r}{0.5\linewidth}
    \begin{center}
    \includegraphics[width=1.1\linewidth]{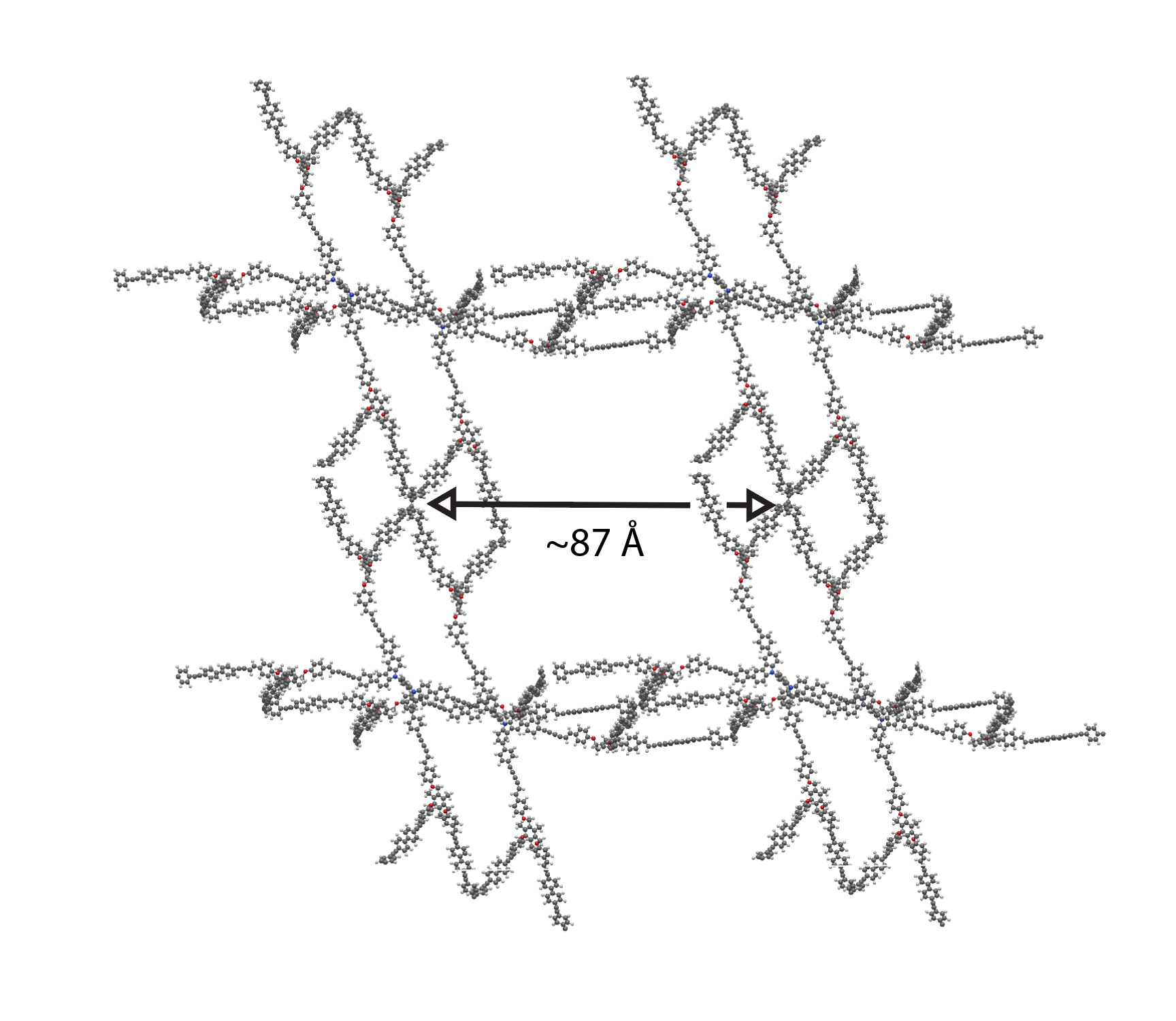}
    \end{center}
    \caption{A render of the relaxed structure of \texttt{005-ffc-10217}, the highest performing structure in the \texttt{matgfn-rm} dataset.}
    \label{fig:best-mof}
\end{wrapfigure}

\textbf{Simulated CO\textsubscript{2} capture performance} In order to confirm the expectation of efficient CO\textsubscript{2} capture from an adsorption proxy (\textit{i.e.}, the gravimetric surface area), we run Physics-based Grand Canonical Monte Carlo simulations for the top-100 generated materials in the \texttt{matgfn-rm} dataset \cite{neumann2022cloud, oliveira2023crafted}. We simulated single-component adsorption isotherms for pure CO\textsubscript{2}, from which we extract the CO\textsubscript{2} working capacity, and dual-component adsorption isotherms for dry flue gas (15\% CO\textsubscript{2} and 85\% N\textsubscript{2}), from which we extract the CO\textsubscript{2}\,/\,N\textsubscript{2} selectivity. All simulations were performed at 300 K, with pressures ranging from 0.15 to 16 bar. The working capacity was calculated as the difference in uptake of (single-component) CO\textsubscript{2} between 16 and 0.15 bar, while the selectivity was calculated as $S = \nicefrac{(Q_{CO_2} / Q_{N_2})}{(f_{CO_2} / f_{N_2})}$, where $Q_i$ is the uptake of species $i$ at 0.15 bar and $f_i$ is the concentration of species $i$ in the input flue gas stream. Figure \ref{fig:adsorption-simulations} shows the distribution of absolute (working capacity) and relative (selectivity) capture metrics for the top-100 \texttt{matgfn-rm} materials. All top-100 materials are (modestly) more selective towards CO\textsubscript{2} than N\textsubscript{2} and exhibit very high CO\textsubscript{2} working capacities, corresponding to the 90\textsuperscript{th} percentile of the experimentally-realised CoRE2019 dataset \cite{Moosavi2020}. Fifteen of the top-100 \texttt{matgfn-rm} materials have working capacities that are higher than all materials found in the CoRE2019 dataset. In particular, we highlight in Figure~\ref{fig:best-mof} the covalent organic framework \texttt{005-ffc-10217} that achieved the highest CO\textsubscript{2} working capacity of the top-100 \texttt{matgfn-rm} materials, around 44 mol/kg. 

\textbf{Relaxation and validity check} Due to the hypothetical nature of the generated MOFs, the crystalline structures are not guaranteed to be perfect. We therefore used the \texttt{mofchecker} library \cite{Mofchecker} to perform basic consistency checks on the generated CIFs. According to \texttt{mofchecker}, all of the top-100 \texttt{matgfn-rm} are porous (metal-)organic materials. However, due to the hypothetical interatomic distances sometimes being larger (or shorter) than the typical bond lengths, some atoms are flagged as either over- or under-coordinated. In order to obtain a more realistic structure, we performed atomic coordinate and unit cell relaxation using the M3GNet \cite{chen2022universal} interatomic potential. Relaxing the structures solves most of the structural problems, with 98\% presenting neither atomic overlaps nor over-coordination of C, N and H atoms, respectively. In particular, for the high-performing \texttt{005-ffc-10217} structure, relaxation led to a 23\% reduction in the unit cell volume, bringing the CO\textsubscript{2} working capacity down to 37.5 mol/kg, which is still larger than those found in the CoRE2019 dataset. The relaxed pore size of \texttt{005-ffc-10217} is approximately 87~\AA.

\section{Conclusion}

In summary, we built a workflow using GFlowNets to generate diverse and novel reticular frameworks with gravimetric surface area greater than 5000 m\textsuperscript{2}/g. As a key result, the top-100 candidates of the resulting \texttt{matgfn-rm} dataset have working capacities in the top 90\textsuperscript{th} percentile of CoRE2019 reference dataset. Moreover, 15 of the top-100 \texttt{matgfn-rm} materials have working capacities that are higher than all materials found in the CoRE2019 dataset. Further tests are underway to confirm the stability and synthesizability of the materials generated in our study. Nevertheless, our results clearly demonstrate  the potential of GFlowNets for materials discovery in carbon capture applications.

\newpage

\section{Contributions}

\textbf{Flaviu Cipcigan}: 
\textbf{Project}: Conceptualisation, Project Administration. \textbf{Paper}: Writing - coordination, Writing – original draft, Writing – review \& editing. \textbf{Software}: Main author of \texttt{matgfn}, contributor to MOF application code, \textbf{Results}: Methodology, Experiments, Formal Analysis, Validation, diversity analysis of \texttt{matgfn-rm} 

\textbf{Jonathan Booth}: 
\textbf{Project}: Project Administration, trained GFlowNets on MOF topologies \textbf{Paper}: Writing – MOF result section, Writing – review \& editing. \textbf{Software}: created reward function and other necessary code for MOF application of GFlownets \textbf{Results}: Methodology, Experiments, Formal Analysis, Validation

\textbf{Rodrigo Neumann Barros Ferreira}: 
\textbf{Paper}: Writing – original draft, Writing – review \& editing. \textbf{Results}: Validation and simulations of \texttt{matgfn-rm} dataset.

\textbf{Carine Ribeiro dos Santos}: 
\textbf{Results}: Validation and simulations of \texttt{matgfn-rm} dataset.

\textbf{Mathias Steiner}: 
\textbf{Project}: Project Administration. \textbf{Paper}: Writing – original draft, Writing – review \& editing.

\section{Acknowledgements}
We thank Dan Cunnington and Won Kyung Lee for valuable comments on the manuscript. Flaviu Cipcigan and Jonathan Booth were supported by the Hartree National Centre for Digital Innovation, a collaboration between STFC and IBM.

\bibliographystyle{unsrtnat}
\bibliography{main}

\newpage
\section*{Appendix A: Training details of GFlowNets on all MOF topologies}

\begin{figure}[H]
    \centering
    \includegraphics[width=1.0\textwidth] {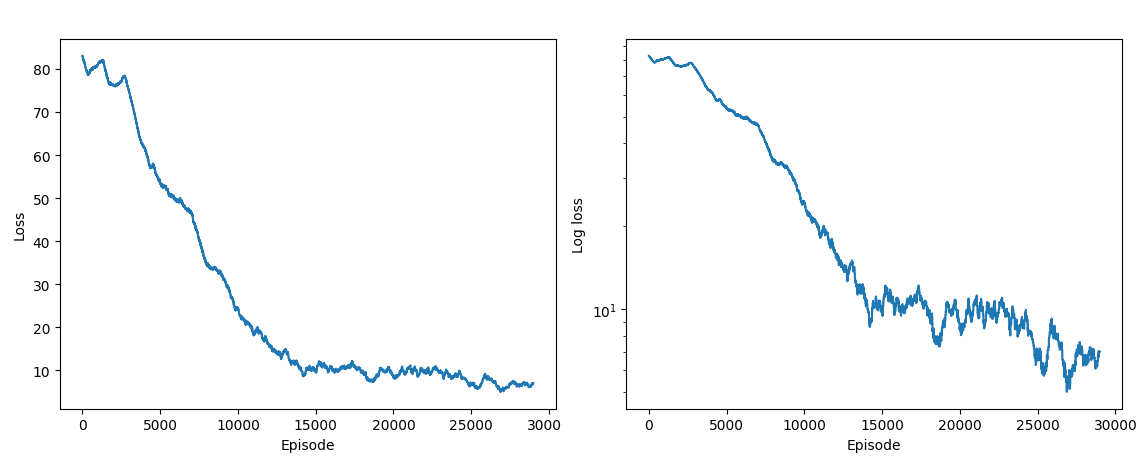}
    \caption{trajectory balance losses for training a GFlowNet on the ASC topology without edges. Losses are smoothed with a 1,000 episode window moving average due to the discovery of a high performing MOF causing a one-episode long spike in the loss.}
    \label{fig:asc_losses}
\end{figure}

\begin{figure}[H]
    \centering
    \includegraphics[width=0.8\textwidth] {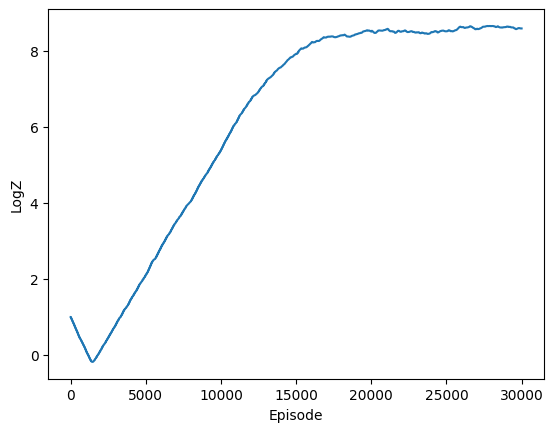}
    \caption{logZ during training for the ASC topology without edges.}
    \label{fig:asc_logZ}
\end{figure}

Figure \ref{fig:asc_losses} shows the trajectory balance losses for training a GFlowNet on the ASC topology without edges while figure \ref{fig:asc_logZ} shows the logZ. All other training runs on other topologies showed similar behaviour. 

The figures below show the performance of the GFlowNet vs random sampling for all eleven topologies with and without edges.

\begin{figure}[H]
    \centering
    \includegraphics[width=1.0\textwidth] {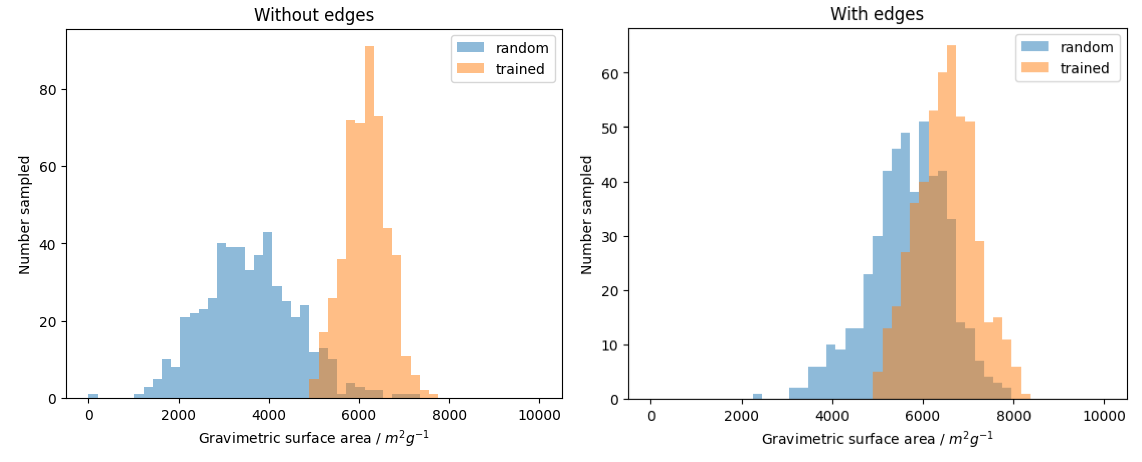}
    \caption{Performance of the GFlowNet trained on the CDZ-E topology.}
    \label{fig:cdz-e_performance}
\end{figure}

\begin{figure}[H]
    \centering
    \includegraphics[width=1.0\textwidth] {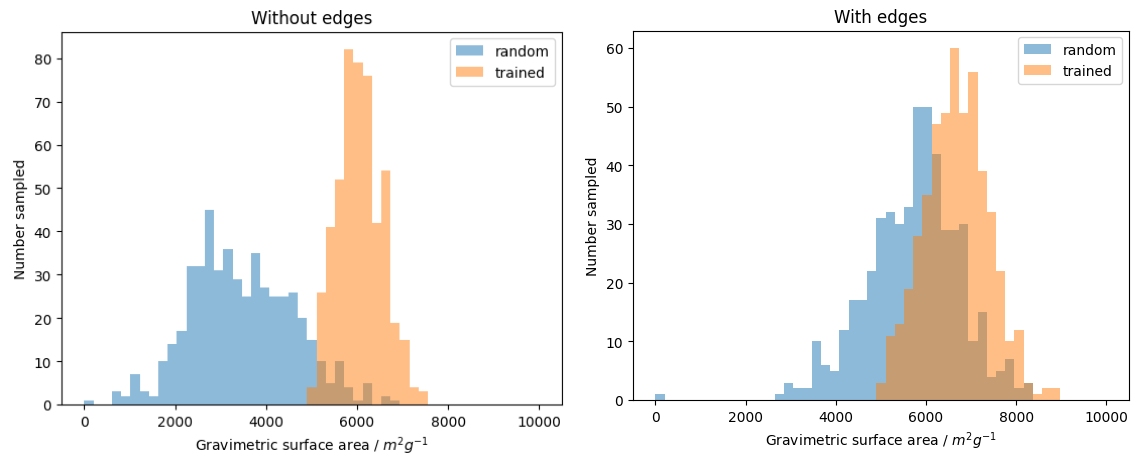}
    \caption{Performance of the GFlowNet trained on the CDL-E topology.}
    \label{fig:cdl-e_performance}
\end{figure}

\begin{figure}[H]
    \centering
    \includegraphics[width=1.0\textwidth] {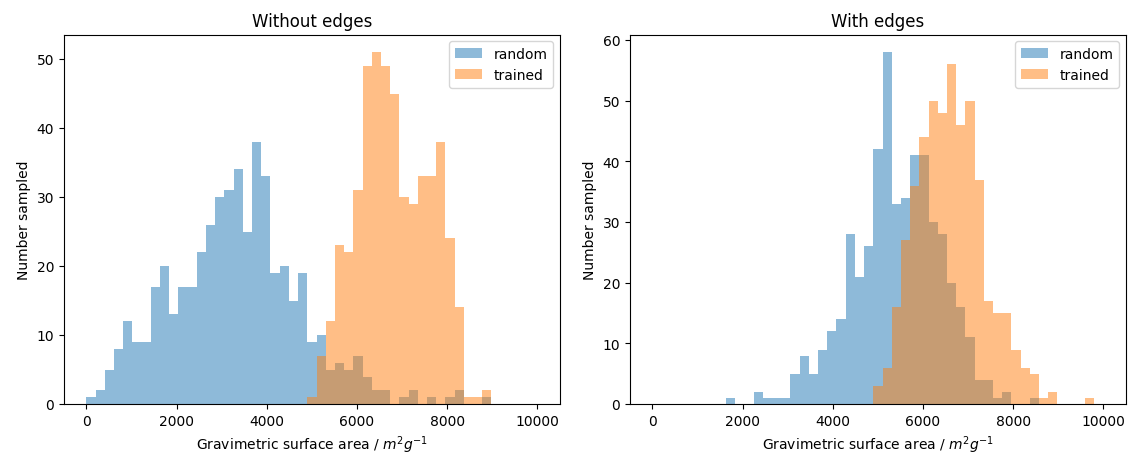}
    \caption{Performance of the GFlowNet trained on the EFT topology.}
    \label{fig:eft_performance}
\end{figure}

\begin{figure}[H]
    \centering
    \includegraphics[width=1.0\textwidth] {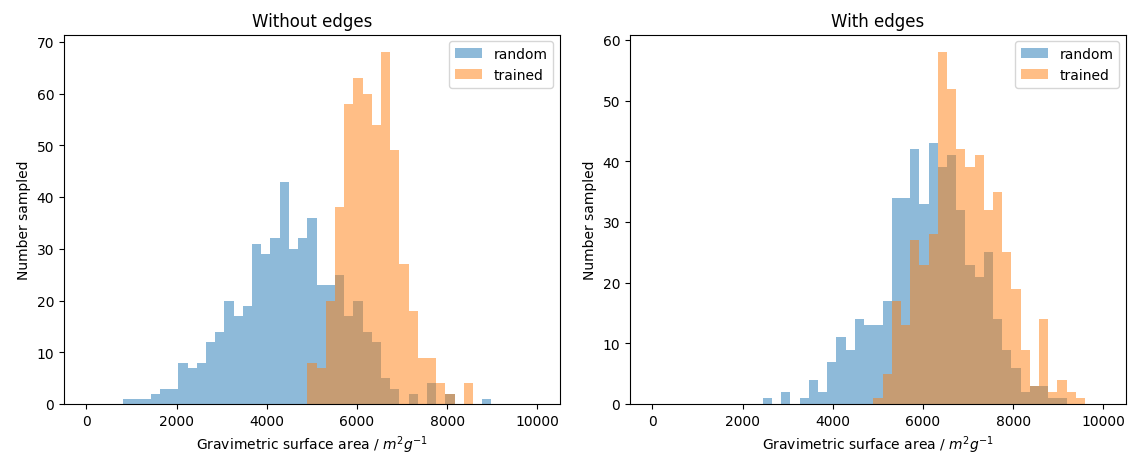}
    \caption{Performance of the GFlowNet trained on the FFC topology.}
    \label{fig:ffc_performance}
\end{figure}

\begin{figure}[H]
    \centering
    \includegraphics[width=1.0\textwidth] {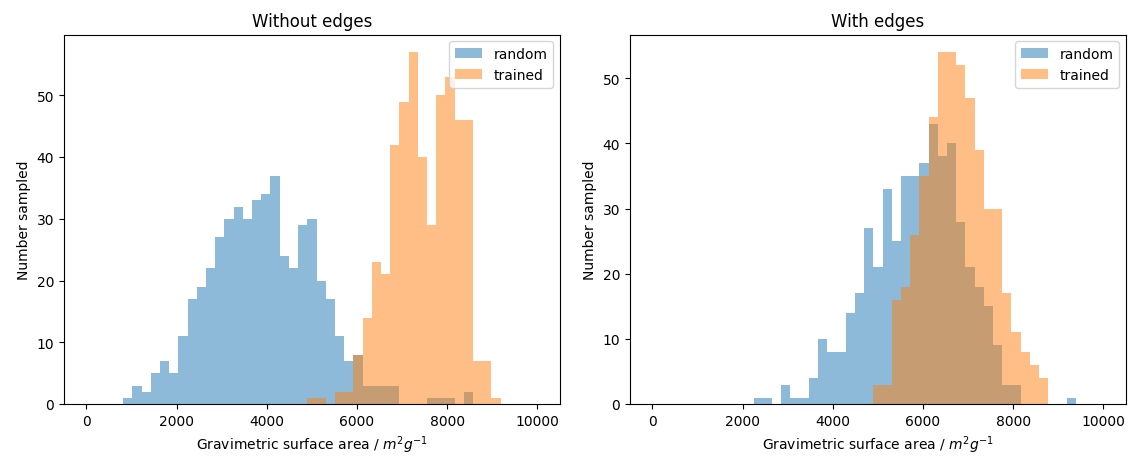}
    \caption{Performance of the GFlowNet trained on the TSG topology.}
    \label{fig:tsg_performance}
\end{figure}

\begin{figure}[H]
    \centering
    \includegraphics[width=1.0\textwidth] {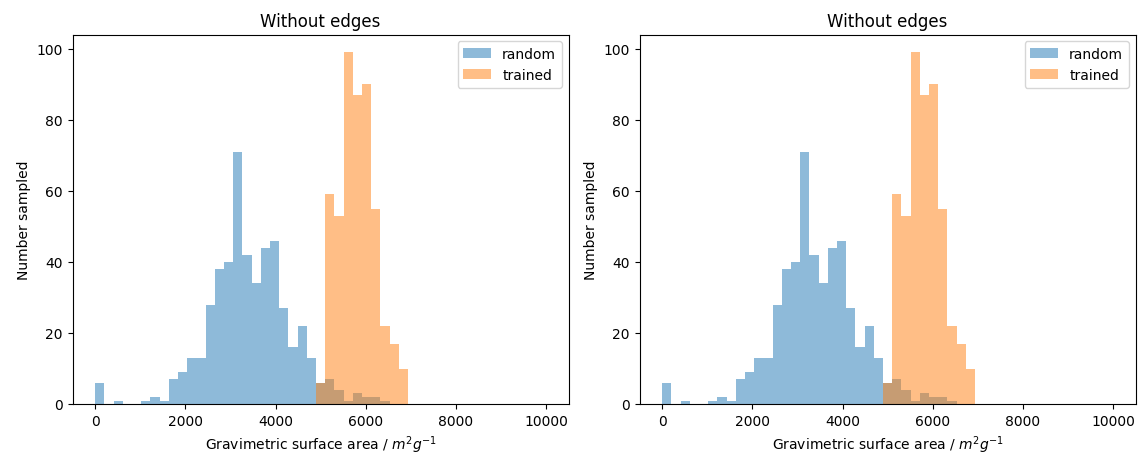}
    \caption{Performance of the GFlowNet trained on the TFF topology.}
    \label{fig:tff_performance}
\end{figure}

\begin{figure}[H]
    \centering
    \includegraphics[width=1.0\textwidth] {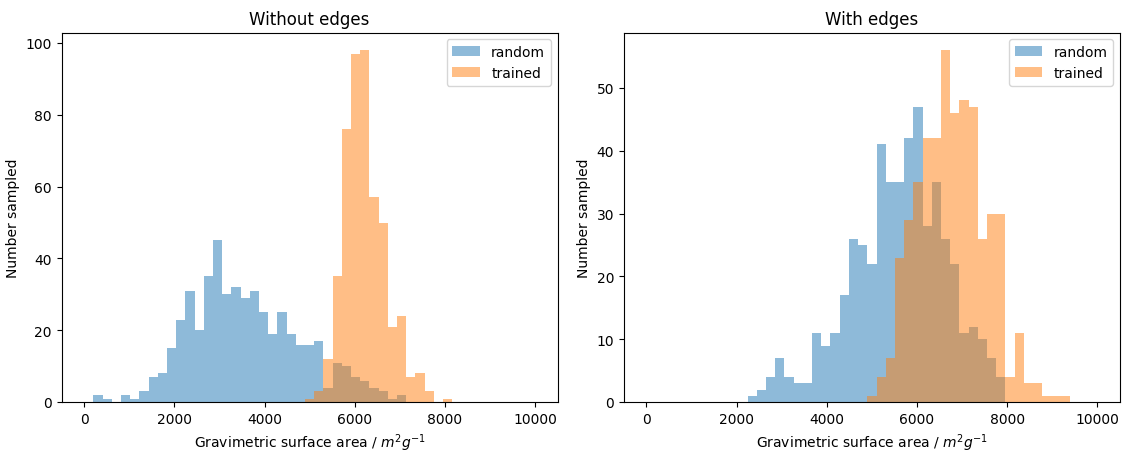}
    \caption{Performance of the GFlowNet trained on the ASC topology.}
    \label{fig:asc_performance_2}
\end{figure}

\begin{figure}[H]
    \centering
    \includegraphics[width=1.0\textwidth] {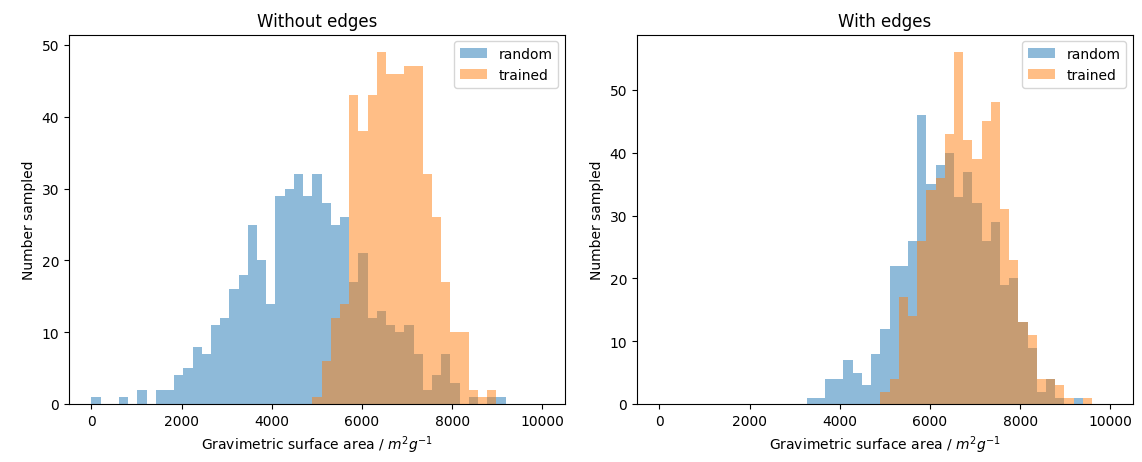}
    \caption{Performance of the GFlowNet trained on the DMG topology.}
    \label{fig:dmg_performance}
\end{figure}

\begin{figure}[H]
    \centering
    \includegraphics[width=1.0\textwidth] {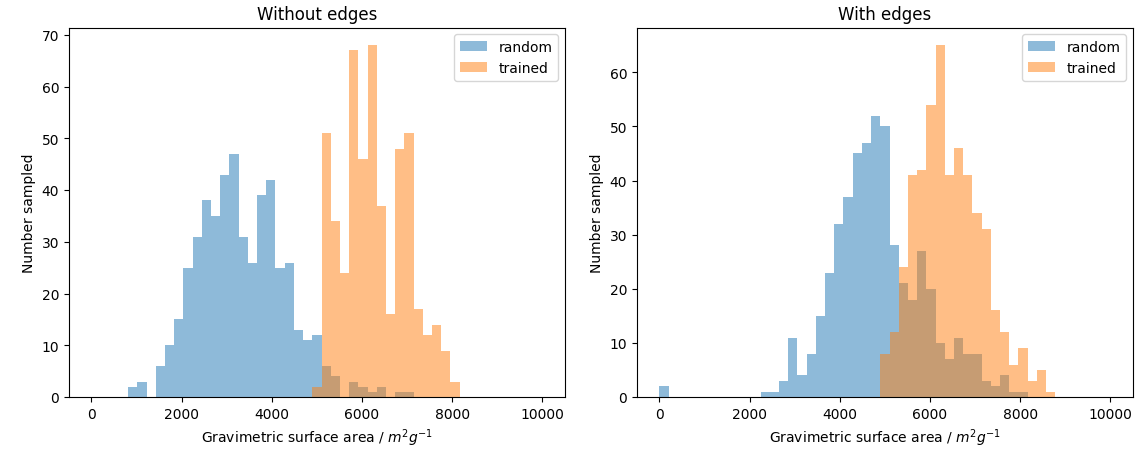}
    \caption{Performance of the GFlowNet trained on the DNQ topology.}
    \label{fig:dnq_performance}
\end{figure}

\begin{figure}[H]
    \centering
    \includegraphics[width=1.0\textwidth] {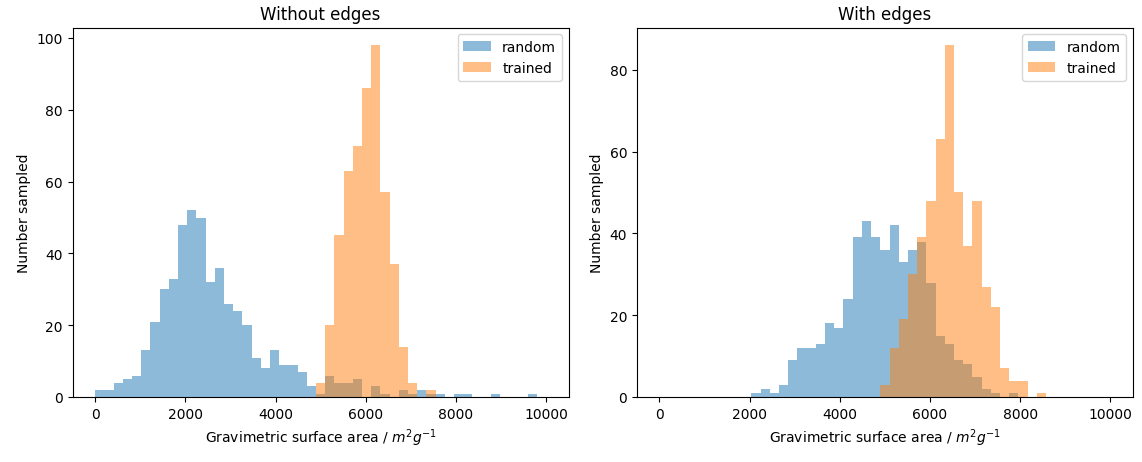}
    \caption{Performance of the GFlowNet trained on the FSO topology.}
    \label{fig:fso_performance}
\end{figure}

\begin{figure}[H]
    \centering
    \includegraphics[width=1.0\textwidth] {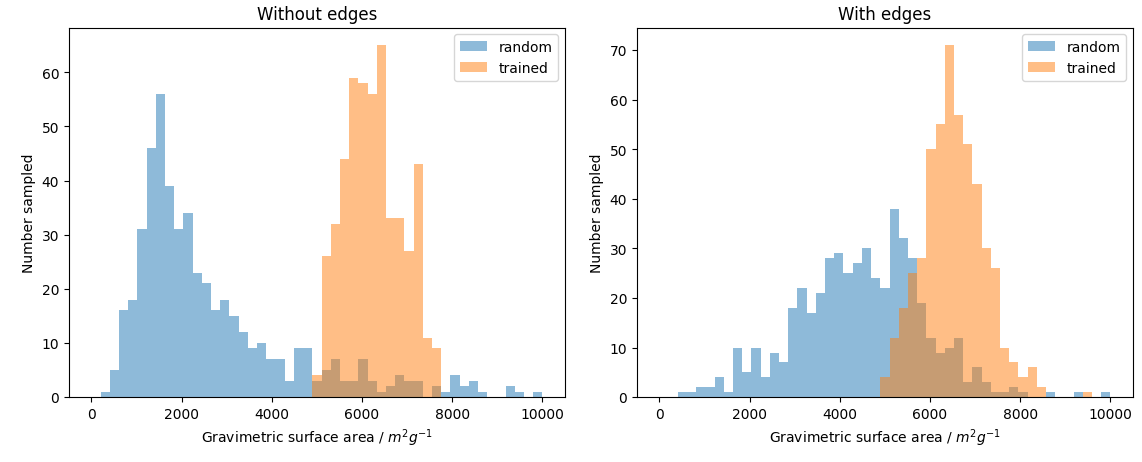}
    \caption{Performance of the GFlowNet trained on the URJ topology.}
    \label{fig:urj_performance}
\end{figure}

\end{document}